\documentclass[11pt]{iopart}
\usepackage{graphicx}
\usepackage{subfigure}
\usepackage{morefloats}
\usepackage{color}
\usepackage{setspace}
\usepackage{floatrow}
\usepackage{amsmath,bm}
\usepackage[pagewise]{lineno}
\begin{document}

\title{Interpretation of scrape-off layer profile evolution and first-wall ion flux statistics on JET using a stochastic framework based on filamentary motion}
\author{ N. R. Walkden$^{1}$, A. Wynn$^{1,2}$, F. Militello$^{1}$, B. Lipschultz$^{2}$, G. Matthews$^{1}$, C.Guillemaut$^{1}$, J. Harrison$^{1}$, D. Moulton$^{1}$ and JET Contributors${*}$
		\\ \small{EUROfusion Consortium, JET, Culham Science Centre, Abingdon, OX14 3DB, UK}
        \\ \small{$^{1}$ CCFE, Culham Science Centre, Abingdon, OX14 3DB, UK} 
        \\ \small{$^{2}$ York Plasma Institute, Department of Physics, University of York, Heslington, York, YO10 5DD, UK} 
		\\ \small{$^{*}$See the author list of “Overview of the JET results in support to ITER” by X. Litaudon et al. to be published in Nuclear Fusion Special issue: overview and summary reports from the 26th Fusion Energy Conference (Kyoto, Japan, 17-22 October 2016)}
        \\ Email: \texttt{nick.walkden@ukaea.uk} }
\date{}

\begin{abstract}
This paper presents the use of a novel modelling technique based around intermittent transport due to filament motion, to interpret experimental profile and fluctuation data in the scrape-off layer (SOL) of JET during the onset and evolution of a density profile shoulder. A baseline case is established, prior to shoulder formation, and the stochastic model is shown to be capable of simultaneously matching the time averaged profile measurement as well as the PDF shape and autocorrelation function from the ion-saturation current time series at the outer wall.  Aspects of the stochastic model are then varied with the aim of producing a profile shoulder with statistical measurements consistent with experiment. This is achieved through a strong localised reduction in the density sink acting on the filaments within the model. The required reduction of the density sink occurs over a highly localised region with the timescale of the density sink increased by a factor of 25. This alone is found to be insufficient to model the expansion and flattening of the shoulder region as the density increases, which requires additional changes within the stochastic model. An example is found which includes both a reduction in the density sink and filament acceleration and provides a consistent match to the experimental data as the shoulder expands, though the uniqueness of this solution can not be guaranteed. Within the context of the stochastic model, this implies that the localised  reduction in the density sink can trigger shoulder formation, but additional physics is required to explain the subsequent evolution of the profile. 
\end{abstract}

\section{Introduction}
The scrape-off layer (SOL) of a tokamak defines the interface between the hot plasma core and cold material surfaces which must be protected. Understanding the properties of the SOL has proven historically difficult due to the complexities of cross-field transport. The transport is known to be highly non-diffusive \cite{GarciaJNM2007,NaulinJNM2007} which makes capturing it in typical 2D transport codes such as SOLPS \cite{SOLPS}, EDGE2D \cite{EDGE2D} or UEDGE \cite{UEDGE} a challenge. In these cases profiles in the upstream SOL are conventionally specified as input with transport coefficients adapted to match. Whilst this approach is of great practical use a first-principles understanding of the SOL cannot be established on this basis, for which a detailed knowledge and decent parameterisation of the transport processes occurring in the upstream SOL are needed. Measurements on many tokamaks worldwide have revealed that SOL transport is intermittent \cite{AntarPoP2003,BoedoPoP2003,GarciaNF2007,DudsonPPCF2005,XuNF2009}, with this intermittency resulting from the radial propagation of coherent structures commonly termed blobs or filaments \cite{D'IppolitoReview}. The propagation of a filament is a complex non-linear problem \cite{KrashenninikovPLA2001} for which numerical simulation is often required \cite{YuPoP2003,GarciaPPCF2006,YuPoP2006}. Whilst the study of filamentary motion is still an ongoing area of active research, many advancements have been made in recent years. These include the role of 3D effects in simple slab geometries \cite{AngusPRL2012,AngusPoP2012,EasyPoP2014} and more complex magnetic geometries \cite{WalkdenPPCF2013,WalkdenNF2015}; the role of finite ion temperature (and associated FLR effects) \cite{MadsenPoP2011,WeisenbergerPoP2014,OlsenPPCF2016,NielsenPPCF2017} and dynamic electron temperature \cite{WalkdenPPCF2016}; the role of electromagnetic effects \cite{LeePoP2015} and magnetic shear \cite{StepanenkoPoP2017}. Indeed the present state of simulations of filamentary motion now permit direct comparison with experiment, as conducted on TORPEX \cite{RivaPPCF2016} and MAST \cite{MilitelloPPCF2016-2}. To properly address the process of SOL formation or evolution, however, many such simulations would be required, or fully saturated turbulence simulations are required \cite{GarciaPoP2005,MilitelloPPCF2013,RicciPPCF2012}. These are extremely computationally expensive and are therefore difficult to run interpretively over parameter spaces required by experiment. It is therefore desirable to have intermediate models that parameterise some of the complexity of the fully non-linear simulations, but are simplified in nature such that they can be effectively `fit' to experimental data. 
\\Garcia \cite{GarciaPRL2012} introduced a statistical characterisation of time-series from single point measurements made in the SOL. This `shot-noise' model treats filaments in the measurement as an uncorrelated train of pulses with an exponential distribution of amplitudes which are ejected in time following a Poisson process. The predictions of the model agree well with measurements on several devices \cite{GarciaPoP2013,GarciaNF2015,GarciaNME2017}, in particular managing to capture the non-Gaussian features of the probability distribution function (PDF) \cite{GravesPPCF2005}. The structure of the PDF is strongly invariant to many changes in plasma conditions \cite{GravesPPCF2005,WalkdenNF2017}, though as shown on JET, this invariance is due to a temporal balance between the duration of bursts and the time between  bursts hitting the probe \cite{WalkdenNF2017}. This stochastic model is limited however in that it deals with measurements made at a single spatial point. Recently Militello and Omotani \cite{MilitelloNF2016,MilitelloPPCF2016} have developed a similar model, based on a statistical description of filamentary motion, to relate spatial profiles in the SOL to the dynamics of filaments. For brevity this model will be referenced herein as the Militello-Omotani (MO) model. The MO model should be viewed as a framework since it is possible to specify a variety of different parameterisations for attributes of the filaments (see the next section for a fuller description) which can then be used to model profiles in the SOL. In such a manner, several possible causes of the experimentally observed non-exponential profiles in the SOL were introduced. Non-exponential profiles, profile flattening or shoulder formation (all used synonymously) describe the tendency of the density profile in the SOL to exhibit distinct regions where the gradient deviates from an exponential profile, often leading to the distinction between a `near' and `far' SOL. This behaviour is remarkably universal across many machines \cite{LipschultzPPCF2002,RudakovNF2005,GarciaNF2007,CarraleroNF2014,MilitelloNF2015,WynnEPS,WynnPaper} and is induced by increasing fuelling rate or reducing the plasma current . Recently these features of the density profile have been linked to changes in the dynamics of filaments on ASDEX-Upgrade \cite{CarraleroNF2014}, which the MO model is ideally suited to investigate.  
\\This paper employs a numerical implementation of the MO model to investigate the process of shoulder formation in the SOL density profile of JET. Using data from the Lithium Beam Emission (LiBES) diagnostic, profiles can be directly compared between the model and experiment. Furthermore time-series data collected at the outer wall by a static Langmuir probe (LP) will be used to compare statistical features of the model with experiment. This paper can therefore be viewed as presenting a methodology for the interpretation of features of the SOL using the MO model with particular emphasis placed on interpreting the formation of the density shoulder.

\section{Experimental reference}
\label{Sec:Exp}
\begin{figure}[htbp]
    \centering
    \includegraphics[width=\textwidth]{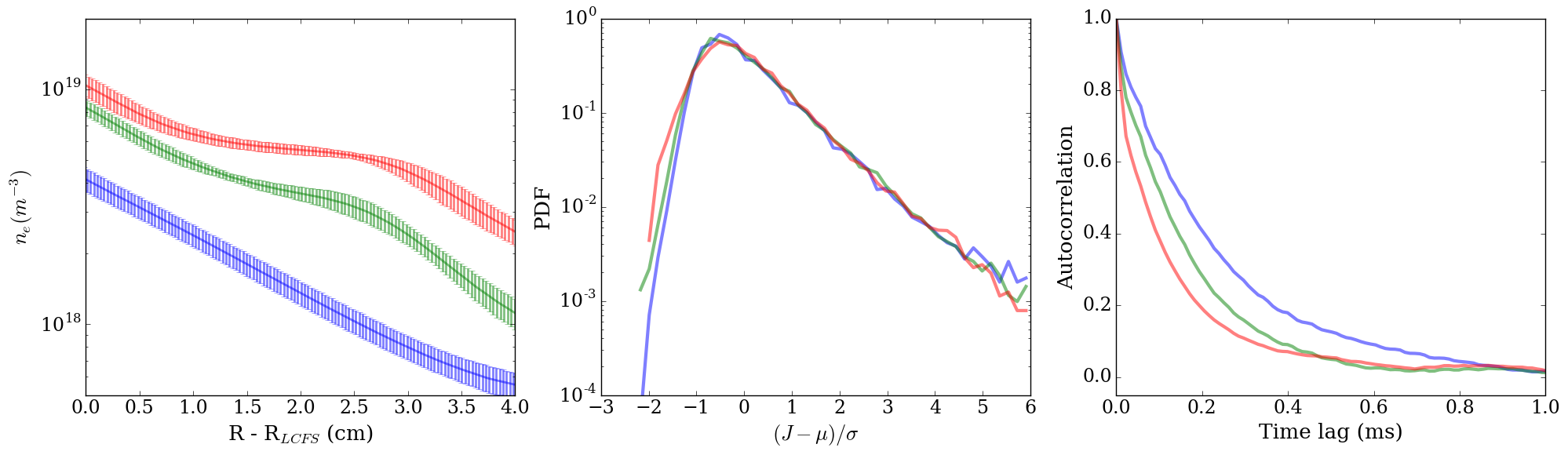}
    \caption{LiBES density profiles during JET pulse 89350 (left) at three different points during the fuelling ramp. Also shown are the PDF shape (centre) and autocorrelation function (right) from a wall-mounted LP. The PDFs and autocorrelation functions are color coded corresponding to the density profile measured at the same time.}
    \label{Fig:Exp_measurements}
\end{figure}

Data taken from JET pulse 89350 will be used as a reference for comparison with the stochastic modelling conducted within this paper. This is an Ohmic L-mode horizontal target plasma with a 2MA plasma current and 2T toroidal field. A fuelling ramp is conducted to systematically increase the separatrix density. The measurements used are density profiles from the LiBES, which is a subset of that presented by Wynn et al \cite{WynnEPS,WynnPaper}, and ion-saturation current time-series from a wall-mounted LP which is a subset of the that presented by Walkden et al \cite{WalkdenNF2017}. In particular the profile shape, PDF shape and autocorrelation function have been measured from three different periods in the density ramp, shown in figure \ref{Fig:Exp_measurements}.
The experimental measurements show a change in the profile structure with the formation of a shoulder in the outer SOL as the separatrix density increases, consistent with previous measurements on JET \cite{CarraleroJNM2015} alongside many other machines \cite{LipschultzPPCF2002,RudakovNF2005,GarciaNF2007,CarraleroNF2014,MilitelloNF2015}. During this flattening of the profile, the PDF shape remains invariant despite a contraction of the autocorrelation function. This can be achieved because the detection rate of filaments on the probe adapts to compensate the shorter duration time \cite{WalkdenNF2017}. It is these three features, the profile shape, PDF shape and autocorrelation function, that will be used for comparison within this paper. For more information on the experimental analysis of the density profile structure see ref \cite{WynnEPS}, and for analysis of the fluctuation characteristics see ref \cite{WalkdenNF2017}.

\section{Stochastic modelling framework}
\label{Sec:Stoch}
\begin{figure}[htbp]
    \centering
    \includegraphics[width=8cm]{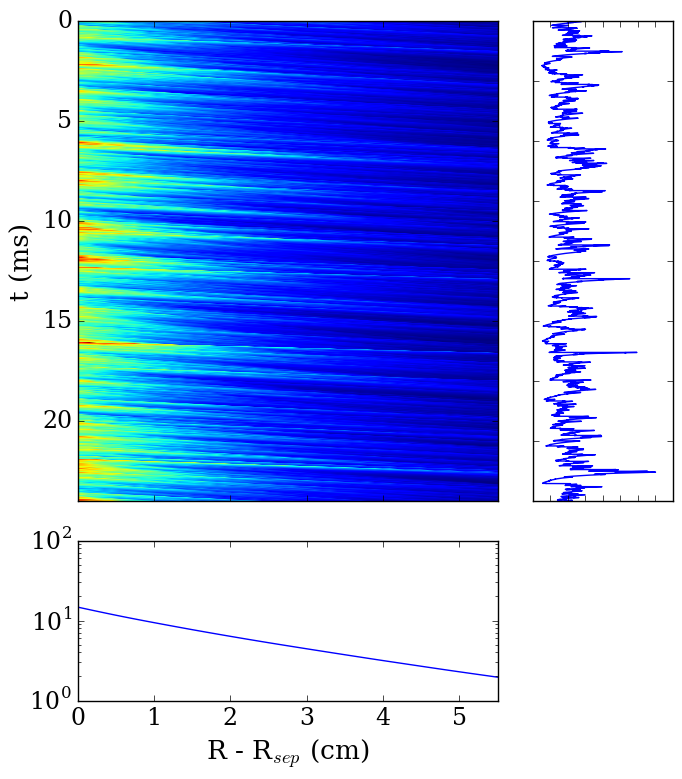}
    \caption{Example of a synthetically produced signal as output from the stochastic modelling described in section \ref{Sec:Stoch}. To the right is an example of the time-series measurement taken at $R-R_{sep} = 5$cm and to the bottom is the time-averaged radial profile measurement.}
    \label{Fig:sigs_schema}
\end{figure}

\begin{figure}[htbp]
    \centering
    \includegraphics[width=16cm]{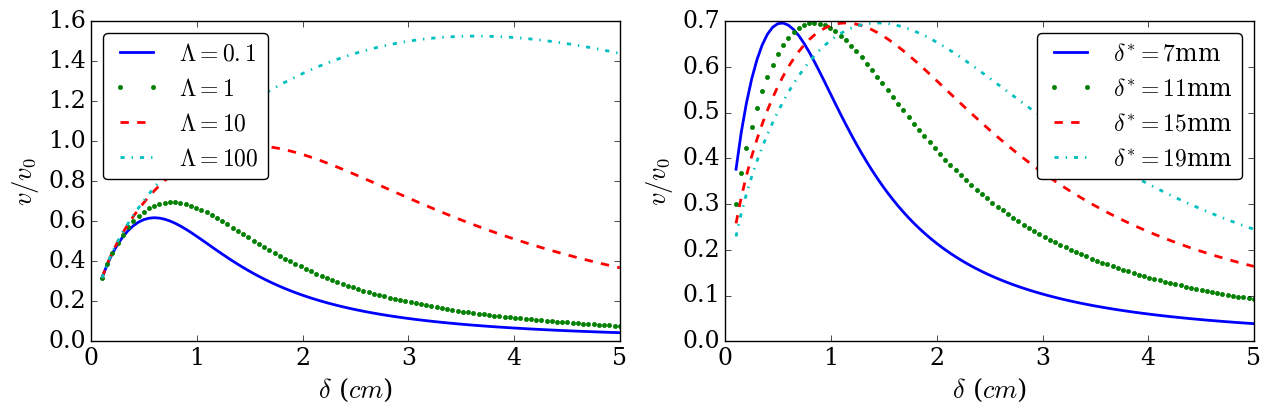}
    \caption{Scaling curves of the normalised filament velocity, $v/v_{0}$, as a function of the filament width $\delta$ across variation in the normalised resistivity $\Lambda$ with $\delta^{*} = 1$cm (left) and critical filament width $\delta^{*}$  with $\Lambda = 1$ (right).}
    \label{Fig:Vels_example}
\end{figure}

In this paper the MO model \cite{MilitelloNF2016,MilitelloPPCF2016} is adopted. A one-dimensional (radial) density signal is constructed by the superposition of a series of statistically distributed pulses with a given spatial waveform propagating radially away from the separatrix with an evolving amplitude. These pulses will be referred to here as filaments. Following Garcia \cite{GarciaPRL2012} the number of filaments ejected in a given interval of time, $\Delta t$, is assumed to behave according to the Poisson distribution such that the the signal at any point along the radial dimension can be described as a `shot noise' process. Physically this means that the ejection of one filament is unaffected by any other ejection event and filaments remain uncorrelated. The ejection rate of the process is given by $f_{f} = N_{f}/\Delta t$ where $N_{f}$ is the total number of filaments ejected with a uniform probability in the time interval given by $\Delta t$. The spatial profile of the i'th filament as defined in ref \cite{MilitelloPPCF2016}, is given by
\begin{equation}
    \eta_{i}\left(x,t\right) = \eta_{0,i}F_{i}\left(x,t\right)\psi\left(x - \int_{0}^{t}v_{i}\left(x,t'\right)dt',\delta_{i}\right)
\end{equation}
where $\eta_{i}\left(x,t\right)$ is the filament radial profile, $\eta_{0,i}$ is the amplitude of the filament density at its ejection time , $F_{i}\left(x,t\right)$ is a function which describes the spatial and temporal evolution of the filament amplitude due to sources/sinks. Here this is characterised by an exponential time-scale such that
\begin{equation}
    F_{i}\left(x,t\right) = \exp\left(\frac{t_{i} - t}{\tau\left(x,t\right)}\right)
\end{equation}
where $t_{i}$ is the ejection time of filament $i$ within the ensemble and $\tau(x,t)$ is the time-scale of the source/sink (negative for a source, positive for a sink) which can in turn depend on spatial position or time. For brevity, $\tau(x,t)$ will be referred to going forwards as the sink timescale. Through this function features such as drainage due to parallel flows or re-fuelling due to ionisation can be introduced.
\\$\psi\left(x,\delta\right)$ is the waveform of the filament and depends on the radial velocity of the filament, $v_{i}\left(x,t\right)$ and the width of the filament, $\delta_{i}$. For all simulations conducted here the filament waveform is taken as a truncated exponential such that $\psi\left(x,\delta\right) = \exp\left(x/\delta\right)H\left(-x\right)$ where $H\left(x\right)$ is the Heaviside function. This shape is motivated by the observation of front formation in filaments as they propagate radially \cite{D'IppolitoReview}. The filament amplitude, $\eta_{0}$, width, $\delta$ and velocity $v\left(x,t\right)$ must be specified and in general here will be drawn from a set of statistical distributions. The spatial and temporal evolution of the velocity and the sink timescale $\tau\left(x,t\right)$ must also be specified. With the required functions and statistical distributions specified a signal can be generated by a summation of the filaments within the ensemble
\begin{equation}
    n\left(x,t\right) = \sum_{i=1}^{N}\eta_{i}\left(x,t-t_{i}\right)
\end{equation}
Here signals will be produced numerically, and measurements of these signals can be made. With the aim of basing this study on a comparison with the experimental observations outlined in section \ref{Sec:Exp}, two aspects of the synthetically produced signals will be measured. Time-averaged profiles will be constructed by averaging each individual radial point over the time frame of the simulation, whilst single point time series will be analysed by taking the signal at each point in time at a single point in radius, here taken at $R - R_{sep} = 5cm$ corresponding approximately to the outer-wall gap for the experiments described in section \ref{Sec:Exp}. Figure \ref{Fig:sigs_schema} provides an example of the synthetic signal produced as a function of radius and time, and the respective time-averaged radial profile and single point time-series at the wall radius.
\\It is well known theoretically that the velocity of a filament has a complex dependancy on its width \cite{YuPoP2003,YuPoP2006,MyraPoP2006,D'IppolitoReview} and its amplitude \cite{GarciaPoP2006,OmotaniPPCF2015} among other parameters. In the `inertial regime' of filament dynamics the filaments radial $\textbf{E}\times\textbf{B}$ velocity is regulated by the ion-polarisation drift and the characteristic filament velocity scales like $v \propto \delta_{\perp}^{1/2}$ where $v$ is the characteristic filament velocity and $\delta_{\perp}$ is the filament width. In the `sheath limited regime' sheath currents regulate the radial velocity which obtains the scaling $v \propto \delta_{\perp}^{-2}$. When resistivity is introduced, the sheath limited scaling is modified to $v \propto (1 + \Lambda)\delta_{\perp}^{-2}$ where $\Lambda$ is the normalised plasma resistivity integrated along the magnetic field line (to be defined later) \cite{MyraPoP2006,EasyPoP2016}. Following the example by Militello and Omotani \cite{MilitelloPPCF2016} a Pade's approximation is used to derive an analytic form for the filament velocity as a function of the filament width
\begin{equation}
    \label{Eqn:vel}
    v = v_{0}\tilde{v}\left(x,t\right)\eta_{0}^{1/2}\frac{\left(\delta/\delta^{*}\right)^{1/2}}{1 + \left(1 + \Lambda\right)^{-1}\left(\delta/\delta^{*}\right)^{5/2}}
\end{equation}
where $v_{0}$ is a scaling parameter for the velocity and $\tilde{v}(x,t)$ contains any spatial and temporal dependance of the filament velocity. Here also the dependancy of the velocity on the filament density amplitude, $\eta_{0}$ has been included. Formally this square-root dependancy holds in the inertial regime, whilst the sheath regime has a more complex dependancy of $v \propto \eta/\left(1 + \beta\eta\right)$ \cite{OmotaniPPCF2015} where $\beta \approx 0.31$ was found to provide a good match to numerical simulation in ref \cite{OmotaniPPCF2015}. Since this more complex dependance does not deviate too significantly from the square-root scaling over a range of a factor $\sim 5$ \cite{OmotaniPPCF2015} it has not been adopted here, however it is possible to include in the Pade's approximation. Finally the parameter $\delta^{*}$ (defined later) determines the transition from the inertial to sheath limited regimes \cite{YuPoP2003}. Figure \ref{Fig:Vels_example} shows the scaling of the normalised velocity, $v/(\tilde{v}v_{0})$ as a function of the width $\delta$ for various values of the control parameters $\delta^{*}$ and $\Lambda$.
\\Finally a choice of statistical distributions must be made for the filament width, $\delta$ and amplitude $\eta_{0}$ at the filament ejection time. For the amplitude distribution function, measurements made with Langmuir probes \cite{GarciaNF2015,WalkdenNF2017} and gas-puff imaging \cite{GarciaJNM2013,GarciaPoP2013} across different machines and conditions show an exponential behaviour. For this reason an exponential amplitude distribution function is adopted here such that 
\begin{equation}
    \label{Eqn:amp_dist}
    P_{\eta_{0}}\left(\eta_{0}\right) = \frac{1}{\eta_{0}^{*}}\exp\left(-\frac{\eta_{0}}{\eta_{0}^{*}}\right)
\end{equation}
For the width distribution function, a log-normal distribution has been chosen. There is less physical basis for this choice due to the inherent difficulties in measuring filament shape experimentally, though distributions of filament widths measured on MAST \cite{AyedPPCf2009,KirkPPCF2016} and NSTX \cite{ZwebenPPCF2016} show a tendency towards a positively skewed asymmetric distribution. Furthermore the choice of a log-normal distribution may be considered appropriate for a positive definite variable such as the width, $\delta$, which may be expected to have a probability that decays to $0$ as the width approaches $0$. As such the distribution of filament widths is taken to be 
\begin{equation}
    \label{Eqn:width_dist}
    P_{\delta}\left(\delta\right) = \frac{1}{\sigma_{\delta}\delta\sqrt{2\pi}}\exp\left(-\frac{\left(\ln\left(\delta\right) - \ln\left(\delta_{0}\right)\right)^{2}}{2\sigma_{\delta}^{2}} \right)
\end{equation}
where $\delta_{0}$ specifies the most probable value of $\delta$ and $\sigma_{\delta}$ is used to specify the width of the distribution.
\\Table \ref{Tbl:Params} tabulates the inputs to the stochastic model employed here that must be set before a simulation can be run. 
\begin{table}[htbp]
    \begin{tabular}{c | l}
        Input & Description \\
        \hline \hline 
        $f_{f}$ & Filament ejection rate \\ 
        $\Delta t$ & Simulation time \\ 
        $ N_{s} $ & Number of samples in $\Delta t$ \\ 
        $\delta^{*}$ & Inertial - Sheath limited transition width in filament velocity relation (\ref{Eqn:vel}) \\ 
        $\Lambda$ & Normalised resistivity parameter in velocity relation (\ref{Eqn:vel}) \\ 
        $v_{0}$ & Velocity scale parameter in relation (\ref{Eqn:vel})\\
        $\delta_{0}$ & Peak width in the width distribution (\ref{Eqn:width_dist}) \\ 
        $\sigma_{\delta}$ & Spread parameter in the width distribution (\ref{Eqn:width_dist}) \\ 
        $\eta_{0}^{*}$ & e-folding amplitude of the amplitude distribution (\ref{Eqn:amp_dist})\\ 
        $\tilde{v}\left(x,t\right)$ & Spatio-temporal evolution function for the filament velocity \\ 
        $\tau\left(x,t\right)$ & Spatio-temporal density sink timescale \\ 
        \hline
    \end{tabular}
    \caption{Table of inputs to the stochastic model used in this paper based off of the Militello-Omotani model.}
    \label{Tbl:Params}
\end{table}
 \\There are a significant number of inputs to the model, some of which can be estimated from experimental conditions, some of which can be constrained by the experimental measurements presented in section \ref{Sec:Exp}, and some of which will be based on assumptions. In the case of $\Delta t$ and $N_{s}$, these are user defined settings which here are set to $N_{s} = 600,000$ and $\Delta t = 1.8s$ which were found to provide statistically converged results. Whilst the profile shape is relatively insensitive to statistical noise, the PDF shape and autocorrelation function are highly sensitive and require long time-series for a suitably accurate result. In appendix \ref{App:Conv} this statistical convergence is demonstrated more clearly.
\\In the next section a baseline simulation will be established to match the experimental measurements early in the density ramps (blue curves in figure \ref{Fig:Exp_measurements}). Sensitivity of this baseline case to assumptions about the inputs will be assessed . From the baseline case, possible mechanisms by which shoulder formation in the SOL profile can occur will be investigated. 
 
\section{Establishing a baseline case}

\begin{figure}[htbp]
    \centering
    \includegraphics[width=\textwidth]{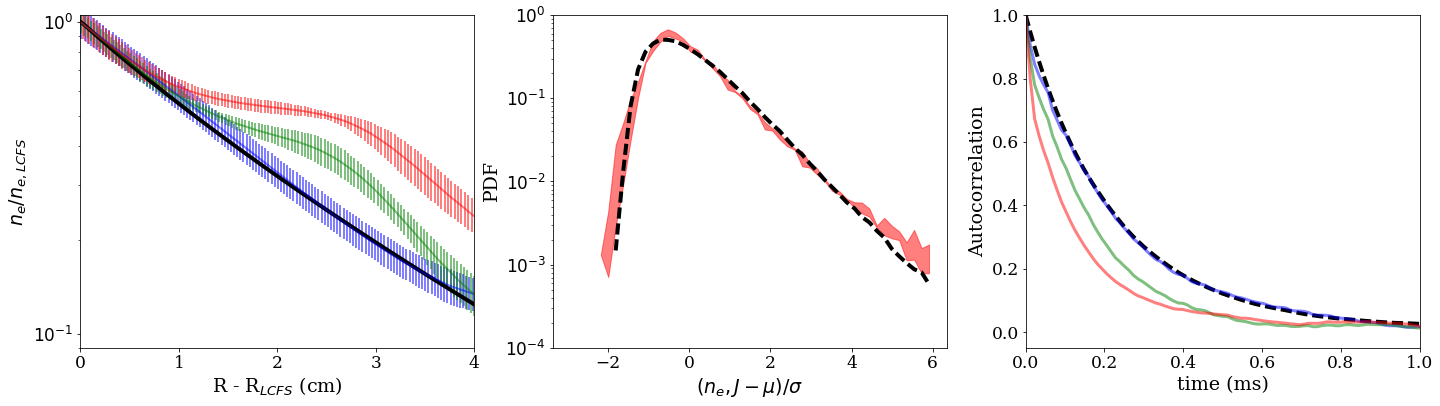}
    \caption{SOL density profile normalised to the separatrix density (left), PDF shape (centre) and autocorrelation function (right) compared between the experimental cases (coloured) and the simulation (black). The experimental data is a repeat of that shown in figure \ref{Fig:Exp_measurements}. In the central plot, the range of the PDF values taken across the full experimental density scan is given, rather than individual measurements, since the PDF shape collapses strongly.}
    \label{Fig:baseline}
\end{figure}

\begin{figure}[htbp]
	\centering
	\includegraphics[width=0.8\textwidth]{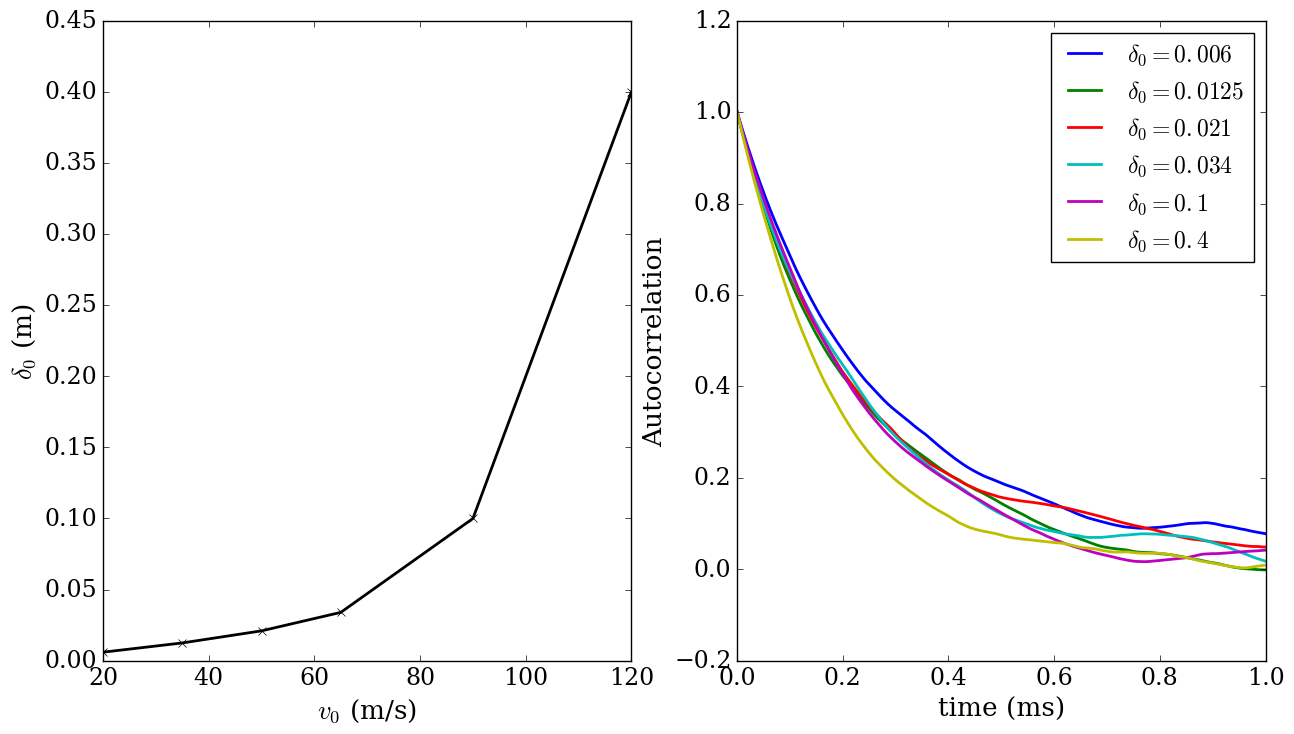}
    \caption{left: Filament width, $\delta_{0}$ required to keep the autocorrelation function approximately fixed as a function of filament velocity, keeping $v_{0}\tau = 0.0285$. Right: Autocorrelation functions for each simulated case showing the approximate similarity in all cases.}
    \label{Fig:Width_scan}
\end{figure}

In this section a baseline simulation will be established to match the characteristics of the experimental measurements at the beginning of the density ramp. This corresponds to the blue data points in figure \ref{Fig:Exp_measurements} where the density profile shows minimal deviation from an exponential. 
\\Several of the inputs in table \ref{Tbl:Params} can be estimated from conditions typical to JET Ohmic L-mode operation. Firstly the sink function is assumed to be dominated by convective losses due to parallel drainage. The sink timescale can therefore be approximated as $\tau = L_{||}/c_{s}$ \cite{Stangeby} where $c_{s} = \sqrt{(T_{e} + T_{i})/m_{i}}$ is the Bohm sound speed. Typical temperatures at the divertor target for JET pulses similar to that used here are in the approximate range $5eV < T_{e} < 30eV$  \cite{WynnEPS,WynnPaper} across the radial profile. The radial increase of $\tau$ resulting from the radial decrease in $T_{e}$ is somewhat compensated by a drop in $L_{||}$ across the SOL. For the JET shot studied here $L_{||}$ drops from 37m in the near SOL at the vertical position of the OLP to 20m in the outer SOL. $\tau$ can therefore be estimated as being within the range $0.48ms < \tau < 0.66ms $ so here it is taken as $\tau = 0.57ms$ across the radial profile. The effect of possible variation in $\tau$ will be covered  later in this section. 
\\Next the parameters $\Lambda$ and $\delta^{*}$ are estimated. These are defined as \cite{YuPoP2003,MyraPoP2006} 
\begin{equation}
    \label{Eqn:dstar}
    \delta^{*} = 2\rho_{s}\left(\frac{L_{||}^{2}}{\rho_{s}R}\right)^{1/5}
\end{equation}
and 
\begin{equation}
    \label{Eqn:Lambda}
    \Lambda = \frac{L_{||}}{c_{s}\tau_{e}}\frac{\Omega_{e}}{\Omega_{i}}
\end{equation}
where $\Omega_{e}$ and $\Omega_{i}$ are the electron and ion gyro-frequencies respectively, $\tau_{e}$ is the electron collision time, $\rho_{s} = c_{s}/\Omega_{i}$ is the Bohm gyro-radius , $R$ is the major radius and $L_{||}$ and $c_{s}$ are defined above. Taking an estimate for the separatrix temperature of $T_{i} = T_{e} = 60$eV give $\delta^{*} = 2.1$cm. The estimation of $\Lambda$ is harder to make since it has such a strong dependancy on the conditions at the divertor target and can easily vary by orders of magnitude. Recent studies on ASDEX-Upgrade \cite{CarraleroNF2014}, JET \cite{WynnEPS} and TCV \cite{VianelloPrePrint} have shown that $\Lambda \approx 0.1$ is typical for the early stages of a density ramp, prior to the formation of the profile shoulder. For this reason $\Lambda = 0.1$ is taken here initially. 
\\Finally, $\eta_{0}^{*} = 1.0$ and $\sigma_{\delta} = 0.3$ are used initially, and both the filament velocity $\tilde{v}$ and the sink timescale $\tau$ are constant in space and time. With these estimations and assumptions in mind an iterative procedure has been followed to optimize the remaining parameters of the model as follows:
\begin{itemize}
    \item First a choice of $\delta_{0}$ is made
    \item $v_{0}$ is varied to match the profile gradient for the given choice of $\delta_{0}$
    \item $f_{f}$ is varied to match the PDF shape for the given $\delta_{0}$ and $v_{0}$
    \item The autocorrelation function is compared and a new choice of $\delta_{0}$ is made
\end{itemize}  
This process is iterated until a good match for the profile gradient, PDF shape and autocorrelation function is achieved. At present this is done manually due to the inherent statistical noise associated with the stochastic modelling making it challenging to automate the process. A good match between the experimental data at the beginning of the density ramp and the simulated measurements is found for $\delta_{0} = \delta^{*} = 2.1$cm, $v_{0} = 50$m/s and $f_{f} = 42$kHz. This is shown in figure \ref{Fig:baseline}.
\\Figure \ref{Fig:baseline} shows that the stochastic model is able to simultaneously reproduce a consistent profile shape, PDF structure and autocorrelation function compared to that of the experiment. Both the filament ejection frequency and the velocity scale factor are found to have values that are in reasonable agreement with recent measurements of filament properties on JET with the ITER-like wall \cite{SilvaNF2014}. Furthermore the value of $\delta_{0}$ here (which should be interpreted as a filament full-width) matches that found by Xu \emph{et al} \cite{XuNF2009} within a reasonable margin, though those measurements were made in inner-wall limited, carbon-wall conditions so are not directly comparable to the measurements considered here. It is worth noting that the value of $v_{0} = 50$m/s obtained here, and effective $\textbf{E}\times\textbf{B}$ velocities observed by Silva \emph{et al} \cite{SilvaNF2014} are somewhat lower than those often found in other devices which typically are on the order of $0.5 - 1$ km/s \cite{D'IppolitoReview}. This may be partly the result of a larger major radius in JET, providing reduced curvature drive compared to smaller sized machines. In terms of the modelling conducted here, $v_{0}$ is determined by the gradient of the Density profile, given a fixed sink timescale. For higher values of $v_{0}$ to be achieved, a stronger sink of the filament density is required which would necessitate larger values of $v_{0}$ to maintain the same profile gradient, however this also impacts the structure of the autocorrelation function. Both a reduction in $\tau$ and an increase in $v_{0}$ lead to a contraction of the autocorrelation function. This contraction can be compensated for by increasing $\delta_{0}$ thus making filaments larger, however this process then becomes naturally limited by the spatial scale of the filaments assuming that it is unreasonable to expect filaments to have spatial scales comparable with the outer-wall gap of the machine (here 5cm which is significantly larger than any direct observation found in literature). In figure \ref{Fig:Width_scan} the value of $\delta_{0}$ required to fit the experimental data is shown as a function of $v_{0}$ holding the product $v_{0}\tau = 0.0285$ fixed to leave the profile gradient unchanged. Also shown is the range of the autocorrelation function produced, showing that it remains approximately fixed and the system is approximately invariant. 
\\As can be seen, with the product $v_{0}\tau$ fixed to maintain the shape of the profile, the filament widths increase non-linearly with the velocity. Both the increase in velocity and the decrease in drainage time (to keep $v_{0}\tau$ constant) have the effect of contracting the autocorrelation function. The only actuator left to counter this effect is the filament width $\delta_{0}$. This means that, if a specific autocorrelation function is sought (as is the case here) then the required increase to $\delta_{0}$ gets larger as filaments move faster, as evidenced in figure \ref{Fig:Width_scan}. Practically, this limits the velocity and sink timescale that can be used as input. Taking here a limit of $\delta_{0} < 5$cm gives $v_{0} < 80$m/s and $\tau > 0.356$ms.  
\\Having established a baseline simulation the next stage of the analysis conducted here is to test whether the stochastic model is capable of reproducing the experimental measurements at the onset of the density profile shoulder, and during its subsequent evolution. Militello and Omotani give three principle means by which non-exponential profiles can be introduced in the stochastic model \cite{MilitelloPPCF2016}. These are:
\begin{enumerate}
    \item{Through changes to the structure of statistical distributions}
    \item{Through changes to the spatial and/or temporal behaviour of the source/sink function}
    \item{Through changes to the spatial and/or temporal behaviour of the filament velocity}
\end{enumerate}
In the next few sections these three different methods will be investigated in turn.

\section{Profile modification within the stochastic model}
In this section the three methods by which a shoulder can be introduced into the SOL density profile will be investigated in turn. It is noted that in principle the different mechanisms may all occur simultaneously in experimental conditions, however this creates a vast parameter space to search manually and is unfeasible here. Instead each mechanism will be investigated in isolation to establish its merit with regards to matching the experimental data.

\subsection{Variation of statistical distributions}
\begin{figure}[htbp]
    \centering
    \includegraphics[width=0.8\textwidth]{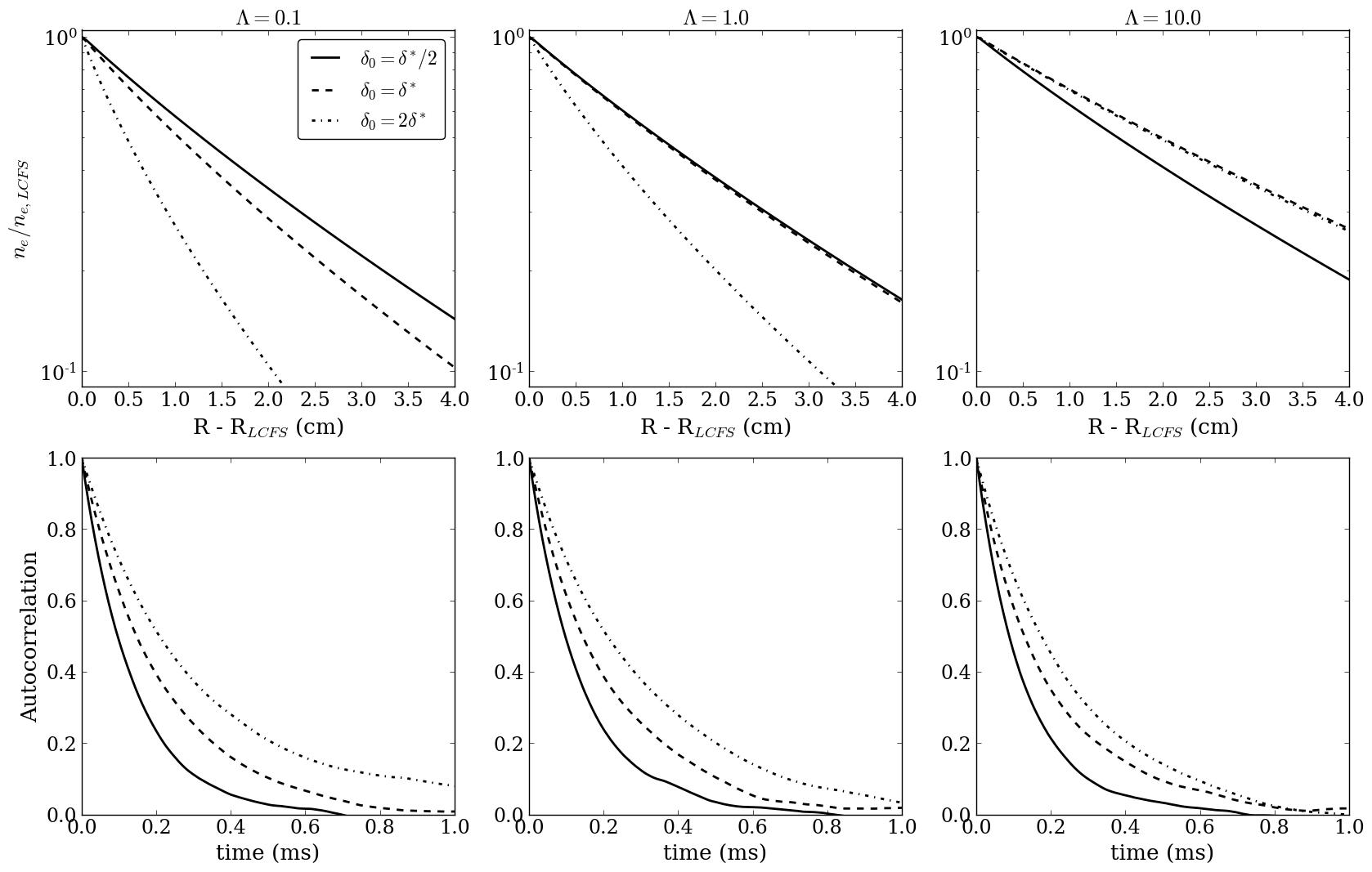}
    \caption{Upper: Normalised profile shape with $\sigma_{\delta} = 0.3$ fixed but varying both $\Lambda$ and $\delta_{0}$ over the ranges given in the text. Lower: Autocorrelation functions measured in the same scenarios as the profiles in the upper row. }
    \label{Fig:Lambda_delta_scan}
\end{figure}

\begin{figure}[htbp]
    \centering
    \includegraphics[width=0.8\textwidth]{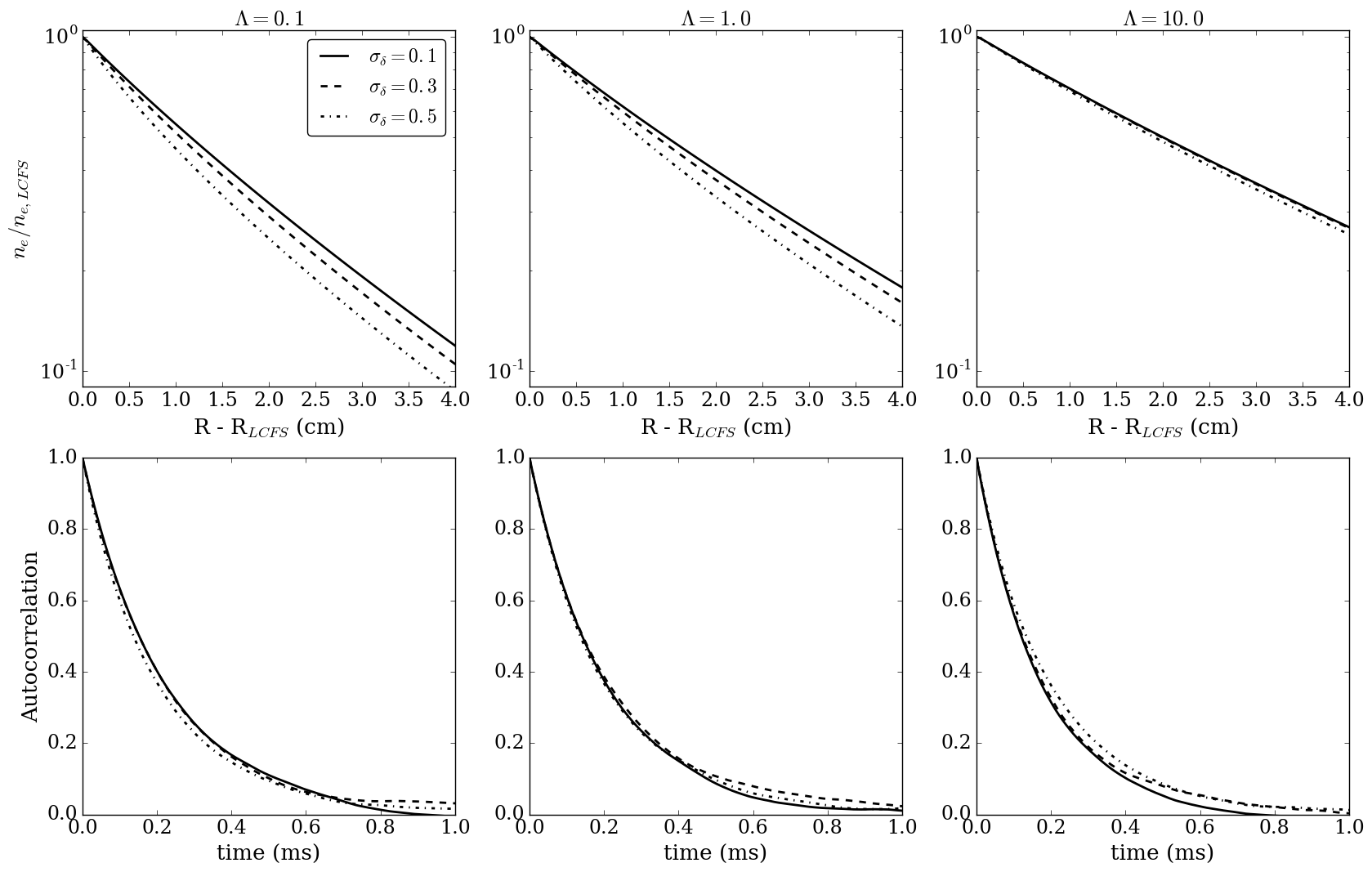}
    \caption{Upper: Normalised profile shape with $\delta_{0} = \delta^{*}$ fixed but varying both $\Lambda$ and $\sigma_{\delta}$ over the ranges given in the text. Lower: Autocorrelation functions measured in the same scenarios as the profiles in the upper row. }
    \label{Fig:Lambda_sigma_scan}
\end{figure}

Through the velocity relation (\ref{Eqn:vel}) both the amplitude and width distributions can impact the distribution of filament velocities, which can in turn impact the mean profile shape. With the amplitude distribution shape fixed as an exponential, and the scaling exponent of the filament velocity on the filament amplitude in equation (\ref{Eqn:vel}) fixed at $0.5$, changes to the e-folding amplitude of the distribution, $\eta_{0}^{*}$, cannot change the profile shape in the manner required to match experiment. To see this note that $v_{0}$ can simply be scaled by $1/\sqrt{\eta_{0}^{*}}$ which then leaves the velocity distribution un-changed. As a result increasing the e-folding amplitude (i.e larger amplitude filaments) is entirely equivalent to scaling $v_{0}$. This may decrease the profile gradient by increasing the velocity of filaments, but it cannot change the shape of the profile in the manner required during the density shoulder formation since any new profiles can always be re-mapped onto the baseline case by a simple scaling of the profile gradient, through $v_{0}$.
\\The width distribution and relationship between the filament velocity and the filament width may be more impactful to the density profile because of the different scalings of the filament velocity in the sheath and inertial regimes. The spread of the width distribution, $\sigma_{\delta}$, as well as the parameters $\delta_{0}$, $\delta^{*}$ and $\Lambda$ may all affect the profile structure via equation (\ref{Eqn:vel}). To test some of these effects, a set of simulations has been run varying both $\Lambda$ and $\delta_{0}$. $\Lambda$ takes the values $\Lambda = 0.1, 1, 10$ which are observed to be approximately the range taken in experiment when shoulder formation and expansion in the density profile is observed \cite{CarraleroNF2014,WynnEPS,WynnPaper}. $\delta_{0}$ values are taken with respect to $\delta^{*}$ as $\delta_{0} = \delta^{*}/2$, $\delta_{0} = \delta^{*}$ (which corresponds to the baseline case at $\Lambda = 0.1$) and $\delta_{0} = 2\delta^{*}$. Figure \ref{Fig:Lambda_delta_scan} shows the profiles and autocorrelation functions calculated for the nine simulations in the scan described here.
\\For the case where $\delta_{0} < \delta^{*}$ variation in $\Lambda$ has minimal impact. This arrises because, as showin in figure \ref{Fig:Vels_example}, below $\delta \sim \delta^{*}$ changing $\Lambda$ has little impact on the filament velocity. Thus only the filaments in the ensemble with widths $ \delta > \sim\delta^{*}$ are significantly impacted by a change in $\Lambda$. It is therefore clear that the width distribution with $\delta_{0} = \delta^{*}/2$ will be the least affected by varying $\Lambda$. Experimentally, $\Lambda$ is often found to be a correlated parameter with the onset of profile shoulder formation \cite{CarraleroNF2014,WynnEPS,VianelloPrePrint} though the causality of the relation remains un-determined. Figure \ref{Fig:Lambda_delta_scan} suggest that if $\Lambda$ is a determining factor in the onset of broadening, then a filament width distribution with a significant fraction of the widths occupying the region $\delta > \delta^{*}$ is required. This should be verified experimentally where $\Lambda$ is thought to be an important parameter. Furthermore $\Lambda$ has very little impact on the autocorrelation function, with significant change only being introduced by a variation in $\delta_{0}$. 
\\The third parameter that influences the width distribution is $\sigma_{\delta}$ which determines the spread in the width distribution. In figure \ref{Fig:Lambda_sigma_scan} the profiles and autocorrelation functions are shown for the three values of $\Lambda$ studied above, this time fixing $\delta_{0} = \delta^{*}$ and varying $\sigma_{\delta} = 0.1, 0.3$ and $0.5$. At $\sigma_{\delta} = 0.5$ the spread of the distribution is beyond what might reasonably be expected from experiment, with a small bit not insignificant proportion of filament widths $\delta > 8$cm. This provides a good test of the sensitivity of the modelling to $\sigma_{\delta}$, which is a relatively unconstrained parameter given the challenge of measuring the filament width distribution experimentally.
\\$\sigma_{\delta}$ has very little impact on the autocorrelation function regardless of the value taken by $\Lambda$. In the profile shape, there is a general trend towards flatter profiles as $\sigma_{\delta}$ decreases, with the importance of this effect decreasing as $\Lambda$ increases. This occurs because, with $\sigma_{\delta}$ taking smaller values, a higher proportion of filaments in the ensemble have $\delta \sim \delta^{*}$ and obtain a larger radial velocity. The effect is not dramatic however and suggests that the results presented here are not sensitive to the value of $\sigma_{\delta}$ within the range $0.1<\sigma_{\delta}<0.5$, which supports the used of $\sigma_{\delta} = 0.3$ in establishing the baseline case. 
\\In both figures \ref{Fig:Lambda_delta_scan} and \ref{Fig:Lambda_sigma_scan} there is a general trend towards flatter profiles and (slightly) contracted autocorrelation functions as $\Lambda$ increases. A notable feature of the profile shape though is a lack of a shoulder region. The profile is slightly sub-exponential and does flatten as $\Lambda$ increases, but does not exhibit the experimental behaviour for any of the values of $\delta_{0}$ and $\sigma_{\delta}$ studied here. Furthermore only a variation in $\delta_{0}$ was able to induce experimentally relevant changes to the autocorrelation function by virtue of reducing the spatial width of the filaments. This suggests that within the stochastic model used here, varying $\Lambda$ or indeed other aspects of the width and amplitude distributions is not sufficient to trigger the formation of the shoulder in the SOL profile. It is worth noting though that varying $\Lambda$ does provide a threshold for changes to the profile and SOL statistics by virtue of the $1 + \Lambda$ factor in the denominator in equation (\ref{Eqn:vel}). The investigation here does not rule out the role of $\Lambda$ or indeed any other of the parameters, but it does suggest that they are not sufficient in isolation and must be accompanied by a change to either the filament velocity or filament drainage. It should also be noted that the parameters tested in this section were held constant across the radial dimension. Allowing variation in these parameters would be a good avenue of future work. This should be treated sensitively however, since it is not presently clear from simulations whether filament motion depends on local conditions spatially, or the conditions present at the formation of the potential dipole that drives the motion. Such a study should be conducted before spatial variation of the parameters in equation (\ref{Eqn:vel}) is introduced. Such a study is beyond the scope of this paper.

\subsection{Filament velocity and drainage}
\begin{figure}[htbp]
    \centering
    \includegraphics[width=\textwidth]{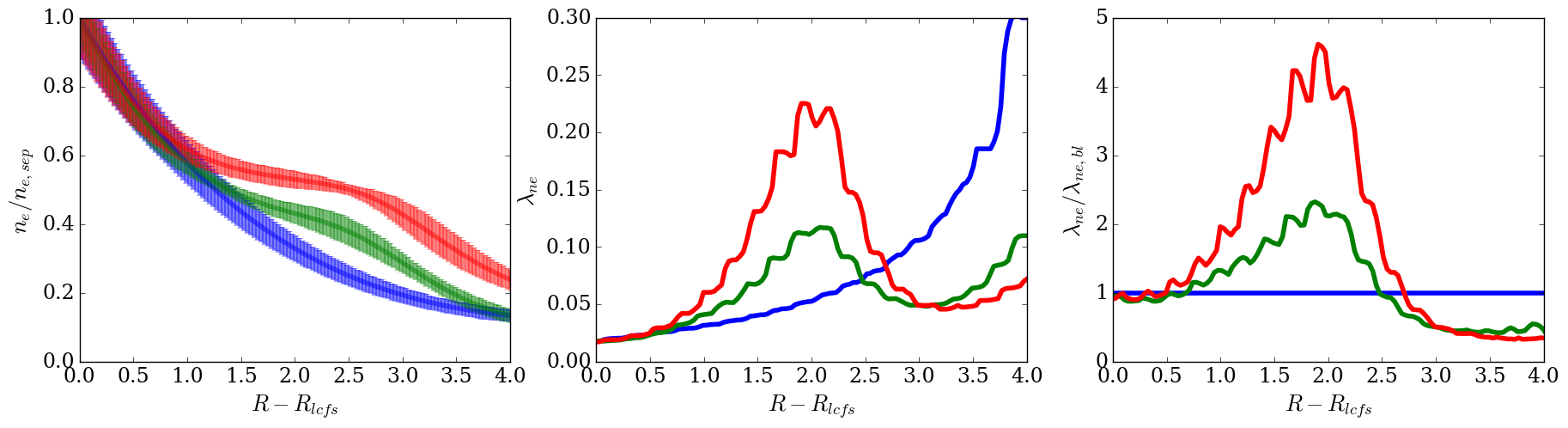}
    \caption{Left: Normalised density profiles showing the onset of profile broadening following the color coding introduced in figure \ref{Fig:Exp_measurements}. Centre: Gradient scale length,$\lambda$, for the density profiles shown to the left. Right: Relative increase observed in $\lambda_{ne}$ compared to the baseline case, $\lambda_{ne,bl}$.}
    \label{Fig:vtau_exp}
\end{figure}

\begin{figure}
	\centering
	\includegraphics[width=0.8\textwidth]{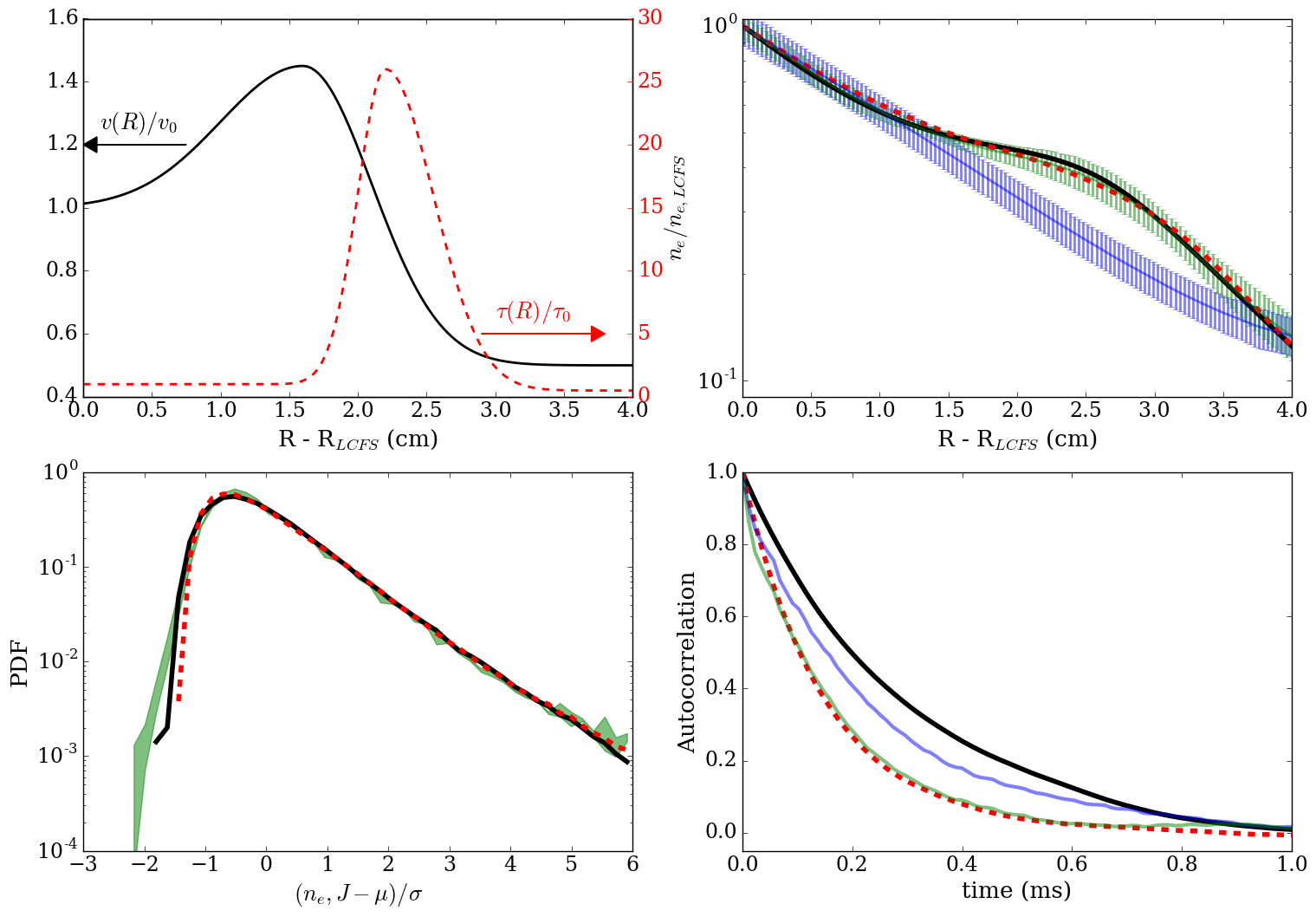}
	\caption{Comparison of the profile shape (upper right), PDF (lower left) and autocorrelation function (lower right) for a modified velocity profile (corresponding to the black curves in all figures) and a modified drainage time (red curves in all figures), which are both shown in the upper right (note the different axes). The profile, PDF range and autocorrelation functions from the baseline case, and the experimental case at the onset of the density shoulder formation are shown for comparison, with colours adhering to the color coding from figure \ref{Fig:Exp_measurements}. }
	\label{Fig:veloc_drain_var_1}
\end{figure}

\begin{figure}[htbp]
	\centering
	\includegraphics[width=0.5\textwidth]{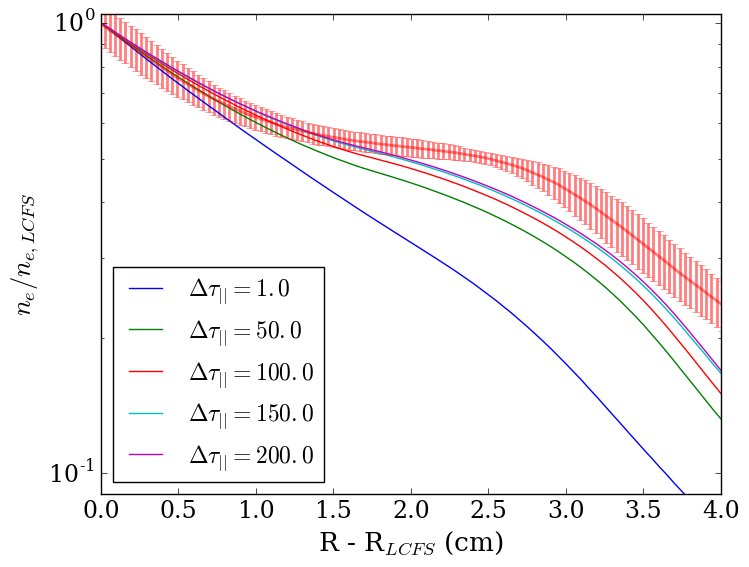}
	\caption{Changes to the simulated profile shape as the degree to which the parallel drainage time is increased with the radially localised profile shown in figure \ref{Fig:veloc_drain_var_1}. The experimental data shown (with errorbars) is the highest density case in the experimental reference shown in figure \ref{Fig:Exp_measurements}.}
	\label{Fig:drain_var}
\end{figure}

\begin{figure}[htbp]
	\centering
	\includegraphics[width=0.8\textwidth]{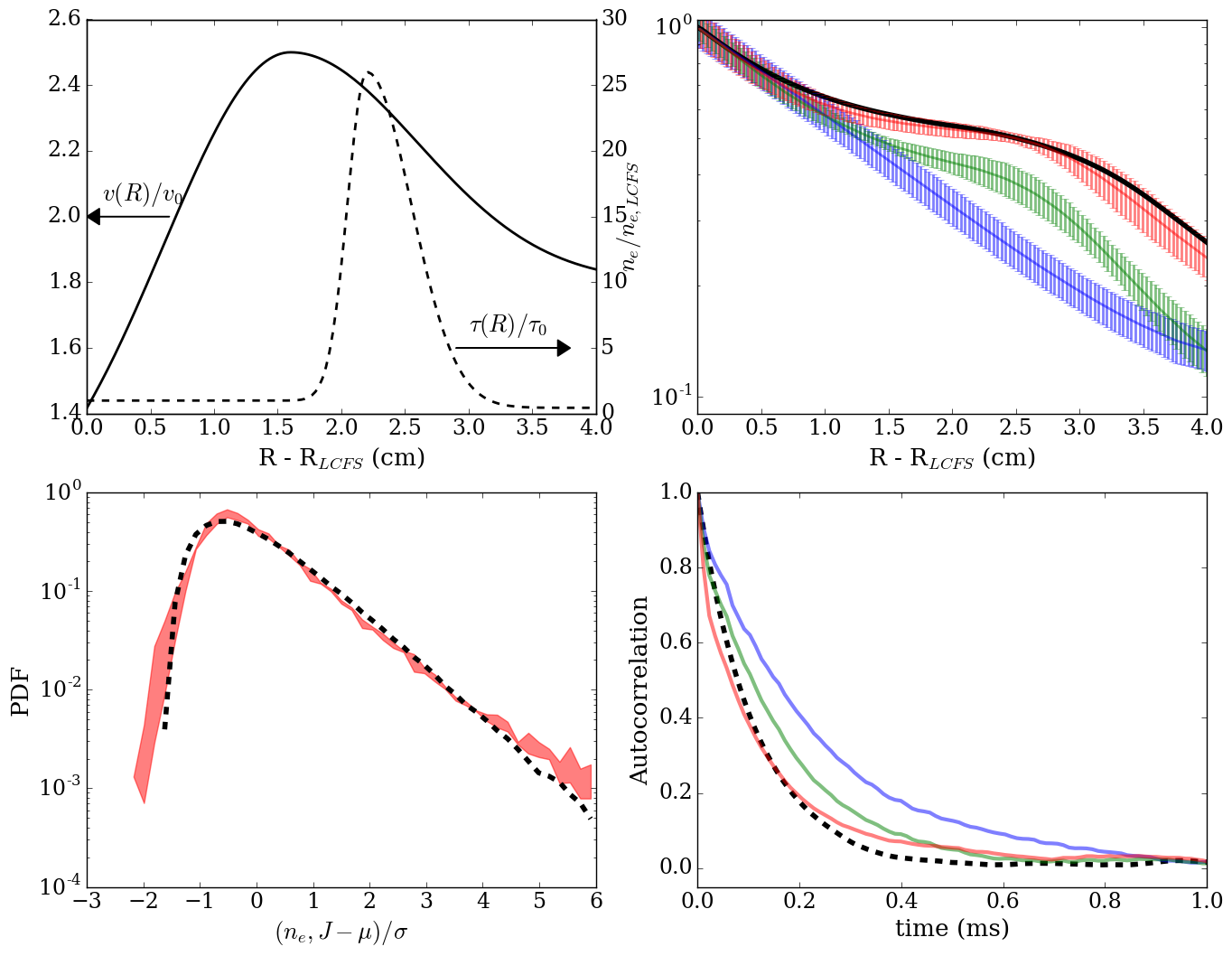}
	\caption{Comparison of the stochastic model with the highest density experimental case allowing for the simultaneous spatial variation of the drainage time and the filament velocity. In the upper left, the profiles of the velocity and density sink (noting the different axes used for each) are shown. In the three comparisons with experiment, black curves indicate the simulated measurements, with the experimental cases following the color coding in figure \ref{Fig:Exp_measurements}. }
	\label{Fig:veloc_drain_var_2}
\end{figure}

It was highlighted in the establishment of the baseline simulation that in the stochastic model used here, as in many studies of scrape-off layer profiles \cite{Stangeby}, the profile shape is set by the competition between parallel and perpendicular transport, which here translates into the variation of $\tilde{v}\tau$ in space and time. The required change in $\tilde{v}\tau$ can be estimated since the local gradient scale length of the profile is approximately $\frac{1}{n}\frac{d n}{dR} = 1/\lambda_{ne} \propto \tilde{v}\tau$ so by evaluating the change in the profile gradient from the baseline case to the broadened profile in the experimental data, the required change to $\tilde{v}\tau$ can be estimated. This is shown in figure \ref{Fig:vtau_exp}.
In the broadened region of the profile an increase in $\lambda \propto \tilde{v}\tau$ by a factor for 2 to 4 is required, whilst in the far SOL a reduction by a half is observed compared to the baseline case. This indicates that either a localised increase in the filament velocity or a localised increase in the density sink timescale is required to capture the shoulder region of the profile. 
\\To parameterise the radial dependance of $\tilde{v}$ and $\tau$ used as input to the stochastic model an asymmetric Gaussian function has been employed such that
\begin{equation}
	g(R) = \begin{cases}
		1 + \Delta_{g}\exp\left(-\left(R - R_{0}\right)^{2}/2\sigma_{g,-}^{2}\right), & \text{for } R \leq R_{0} \\
		g_{0} + \left(\Delta_{g} + 1 - g_{0}\right)\exp\left(-\left(R - R_{0}\right)^{2}/2\sigma_{g,+}^{2}\right), & \text{for } R > R_{0} 
	\end{cases}	
\end{equation}
where $g$ represents either the sink timescale or the velocity and $\sigma_{g,-} \neq \sigma_{g,+}$. Once again a manual optimisation of the velocity and drainage function parameters has been used, such that the results may be considered reasonable but not fully optimised. Starting from the baseline case the velocity function and the sink timescale have been varied respectively in separate simulations to try and match the profile shape at the onset of the shoulder formation (green data points in figure \ref{Fig:Exp_measurements}). Figure \ref{Fig:veloc_drain_var_1} shows the comparison between the profile structure, as well as the PDF and autocorrelation functions for the cases where the velocity and drainage time have been adapted respectively in separate simulations.
\\Figure \ref{Fig:veloc_drain_var_1} shows that within the stochastic model used here it is possible to induce an experimentally consistent shoulder in the density profile through either a localised increase in the filament velocity or in the density sink timescale. The peaks of the $\tilde{v}$ and $\tau$ are offset from one another. This is the result of non-local effects that occur due to the spatial structure of the filaments. When filaments are accelerated they encounter a less severe drainage so transport a higher density into the outer SOL, raising the profile in that region. However filaments in the acceleration zone become more rarified such that on time-averaging, the density in that region can be reduced. The exact opposite effects are true in the case of deceleration, as is present in the outer SOL in figure \ref{Fig:veloc_drain_var_1}. The density sink on the other hand has a spatially localised affect on the filaments since their loss rate is determined at each point in the profile by the sink at that point. It is also notable that in order to match the experimental profile by changing the sink timescale, an increase by a factor 25 is required. If the density sink is taken as the result of parallel losses, this implies a near stagnation of any flows that are providing this loss. Alternatively this may imply a refuelling of the filament via, for example, ionisation processes. The stochastic model is not able to determine the physical mechanism at play beyond noting that it impacts the density sink. 
\\Although both filament acceleration and an increase in the density loss timescale give reasonable profile shapes and leave the PDF shape invariant, the autocorrelation function serves to distinguish the two cases. In the outer part of the density profile, the gradient increases outside of the shoulder. To capture this increase, either a deceleration of the filament velocity or a reduction in the loss timescale is required. These have opposing effects on the autocorrelation function, with the former leading to an expansion and the latter leading to a contraction.  Comparing this important difference with the experimental case suggests that the localised acceleration/deceleration of filaments is not consistent with experiment, whilst the localised increase in the sink timescale does produce consistent results. Indeed the value of $\tau$ that produces consistent profile behaviour also produces a consistent reduction in the autocorrelation time without optimising any other parameters (filament width for example). Thus in the context of the stochastic model used in this paper, a radially localised increase in the density sink timescale is the simplest (in the sense that only a single parameter is varied) method by which a profile shoulder can be triggered whilst maintaining experimentally consistent statistics at the outer-wall. 
\\It is not a simple matter to extend the comparison based on variation of the density sink timescale to the highest density case in the experimental reference (the red data traces in figure \ref{Fig:Exp_measurements}). Figure \ref{Fig:drain_var} shows the result of sequentially increasing the parameter $\Delta_{\tau}$ which determines the level to which the $\tau$ is increased, compared to the high density experimental profile. 
As $\tau$ is increased the level to which the profile is affected lessens and the experimental profile cannot be reproduced. Furthermore, and importantly, because the profile gradient in the outer-SOL is moderately decreased in the high density case, the sink timescale cannot be altered in a manner that simultaneously captures the gradient decrease \emph{and} the contraction of the autocorrelation function. This implies that additional physics is required to capture the subsequent evolution of the profile following shoulder formation. It has been possible to generate a stochastic signal that gives reasonable agreement with the experimental data after the shoulder formation by allowing for simultaneous variation of the sink timescale and velocity profiles. This is shown in figure \ref{Fig:veloc_drain_var_2} where spatial variation in both $\tilde{v}$ and $\tau$ are included in the same simulation.
A mild acceleration of the filament velocity alongside the strong increase in the loss timescale couple to provide conditions that match both the profile shape and autocorrelation function. This is a plausible manner in which the profile evolution can be captured within the stochastic model, but an exhaustive search has not been conducted. What is clear from this study is that there is an additional physics requirement to capture the profile evolution, but that the stochastic model does have the capability to do this consistently with experiment.

\section{Discussion}
\begin{figure}[htpb]
    \centering
    \includegraphics[width=1.2\textwidth]{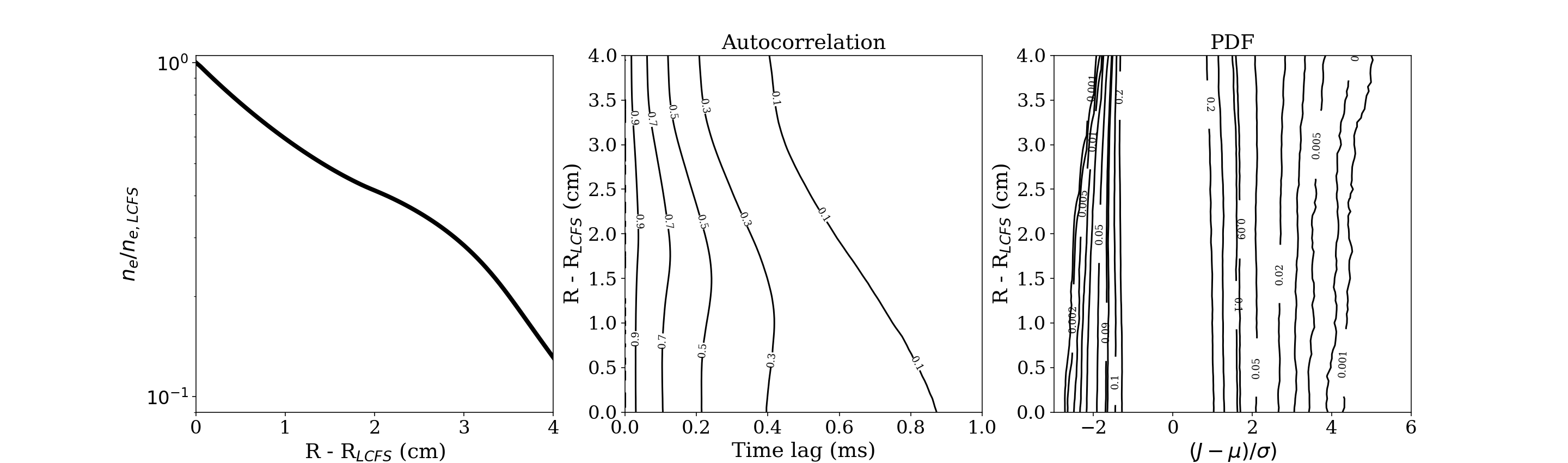}
    \caption{Left: Radial profile taken from the case at the onset of shoulder formation, shown in figure \ref{Fig:veloc_drain_var_1}, using a modification of the sink function. Centre: Contours of the autocorrelation as a function of the major radius. Right: Contours of the PDF as a function of the major radius.}
    \label{Fig:AutoCorr_Radial}
\end{figure}

The comparison made with the experimental data on JET, presented here, shows that the stochastic model is capable of reproducing consistent statistical measurements as well as profile measurements at various stages of the profile evolution. Whilst there are certainly limitations to this modelling approach, this level of agreement is encouraging and should motivate wider use of stochastic modelling applied to SOL phenomena. Of particular note is the observation here that a drastic increase in the density loss timescale was found the be the simplest (in terms of requiring no other changes to the model) consistent way of matching the change in profile shape and change in autocorrelation function at the onset of the profile shoulder formation. This comparison was based on probe measurements made at the wall radius where the auto-correlation function contracts as the profile shoulder forms. In the centre of the profile, where the gradient relaxes as the shoulder forms and the loss timescale is drastically increased and the opposite trend is evident. Indeed the variation of the density sink radially leads to a radial variation in the autocorrelation function. This is shown in figure \ref{Fig:AutoCorr_Radial} which shows the autocorrelation function and PDF shape as a function of radius at the onset of the shoulder formation.
\\In the centre of the profile, where the shoulder forms, the autocorrelation function is at its widest since the density loss timescale as longest, whilst the PDF shape becomes increasingly skewed further away from the separatrix. Measurements made on ASDEX-Upgrade \cite{CarraleroNF2014} and TCV \cite{VianelloPrePrint} have shown such an increase in the autocorrelation time within the shoulder of the profile. A good test of this theory may also be available on MAST, where profile shoulder formation was observed \cite{MilitelloNF2015} without the secondary increase in the profile gradient in the far SOL. In this case, if the filament drainage is the responsible actuator for the profile shoulder formation then a probe situated anywhere in the far SOL should measure a widening of the autocorrelation function as opposed to the contraction measured on JET. 
\\The physical mechanism that may lead to such a strong increase in the loss timescale is not something that the stochastic modelling can determine. Militello and Omotani have suggested that strong momentum loss along the magnetic field line due to charge exchange collisions with neutral atoms in the divertor \cite{MilitelloPPCF2016} may be a possible cause. Alternatively strong ionisation in the divertor may induce a reduction and eventual reversal of parallel flows. Notably this would imply a dependancy on the geometry of the divertor and in particular the poloidal orientation of the divertor plate. Such a dependancy has been noted on JET where plasmas run in the vertical target (VT) configuration showed a lack of profile shoulder formation compared to those in the horizontal target (HT) configuration \cite{WynnEPS,WynnPaper}. Another implied dependancy is on the neutral density. Again in JET plasmas with similar divertor conditions driven by Deuterium seeding and Nitrogen seeding respectively, plasmas with N-seeding were significantly more resistant to the shoulder formation than plasmas with D-seeding. There is therefore an evidence base to support the assertion that density sink function may be responsible for shoulder formation. Furthermore strong D$_{\alpha}$ radiation in the divertor of JET has been postulated to be associated with the presence of a shoulder in the SOL density profile \cite{WynnPaper}, however it is not clear from experiment what role is being played. It is also worthwhile noting that the stochastic modelling suggests that an increase in the loss time can trigger shoulder formation, but cannot capture the subsequent profile evolution in isolation. This suggests the evolution of the shoulder and the formation of the shoulder are related, but different events and should possibly be viewed separately from one another. One such mechanism was found to be a radial change in the filament velocity. One possible trigger for such a change could the profile of $\Lambda$ measured at the divertor. In this paper $\Lambda$ was kept constant radially, however in experiment $\Lambda$ is observed to vary from the separatrix outwards and may provide a mechanism for the widening of the shoulder region, as asserted by Carralero \emph{et al}\cite{CarraleroNF2014}. Such a variation in $\Lambda$ could be included in the stochastic model here, and would be a good avenue of future work.

\section{Conclusions}
This paper has presented a first attempt at interpreting experimental scrape-off layer density profiles and ion-flux statistics to the outer wall in JET using a stochastic model based on filament dynamics. The stochastic model is used to match measurements of the profile simultaneously with measurements of the autocorrelation function and PDF shape of the ion-saturation current measured at the outer-wall. It treats filaments stochastically with an amplitude, ejection time and width drawn from statistical distributions and with a velocity determined by a theoretically motivated relationship with the filament width. By establishing a baseline case prior to the onset of density profile shoulder formation, the stochastic model has been manually optimised to match the experimental measurements. The degree of agreement between the model and the measurements is encouraging and shows that with sufficient inputs, the model is capable of simultaneously capturing the time-averaged profile and the spatially localised statistical measurements. A sensitive study was carried out to assess how the model varied as some of the poorly constrained input parameters were varied. In particular, with the product of the velocity and the timescale of the density sink of filaments fixed to maintain a fixed profile gradient, the filament width was varied to keep the autocorrelation function fixed as the velocity was increased. The required filament width grew exponentially and placed reasonable strict limits on the possible velocities that could be used as input without requiring unrealistically large filaments.
\\With the baseline case established, inputs to the stochastic model were varied to recreate the formation of a shoulder in the density profile alongside a contraction of the autocorrelation function. Changes to the input statistical distributions and parameters associated with the filement velocity-width relationship were shown to be unable to capture the change in the profile shape on their own. By employing a spatial variation in either the filament velocity or the loss timescale of the density, the modified profile shape was captured. The contraction of the auto-correlation function was then used to eliminate the spatial variation of velocity since this lead to an expansion which opposed the experimental measurement, whilst the loss time variation lead to the correct degree of contraction in the autocorrelation function without further optimisation. Within the stochastic modelling framework this suggests that a strong increase in the density sink timescale is the simplest way to trigger the onset of density profile shoulder formation. The required increase in the loss time is strongly spatially localised in the region of shoulder formation and represents a factor 25 increase over the baseline level. 
\\It was not possible to match the profile structure later in the density scan, as the shoulder region widened and flattened, with just an increase in the filament density loss time. This implies that, within the stochastic model, shoulder formation can be triggered by the increase density loss timescale, but the subsequent evolution of the profile requires additional physics. Including in addition a radial variation in the filament velocity alongside the localised increase in the sink timescale was shown to be sufficient to match the profile structure, PDF shape and autocorrelation function at the higher density case. The uniqueness of this solution cannot be guaranteed, but this does imply that additional physics, beyond an increase in the density sink, is required for an expansion of the density shoulder in the stochastic model. 
\\The stochastic model is shown here to be a potentially useful tool allowing for analysis of physics in the SOL on the basis on non-local intermittent transport without the demands of non-linear fluid codes. It can recreate profile shapes and consistent statistical measurements with appropriate inputs. Whilst there is certainly more development of the method required, it is hoped that this motivates its use going forwards. 

\section{Acknowledgements}
This work has been carried out within the framework of the EUROfusion Consortium and has received funding from the Euratom research and training programme 2014-2018 under grant agreement No 633053 and from the RCUK Energy Programme [grant number EP/I501045]. To obtain further information on the data and models underlying this paper please contact PublicationsManager@ukaea.uk. The views and opinions expressed herein do not necessarily reflect those of the European Commission

\section{References}
\bibliographystyle{prsty}
\bibliography{../Bibliography}

\appendix\section{Appendix: Statistical convergence of stochastic modelling signals}
\label{App:Conv}
\begin{figure}[htbp]
    \centering
    \includegraphics[width=0.8\textwidth]{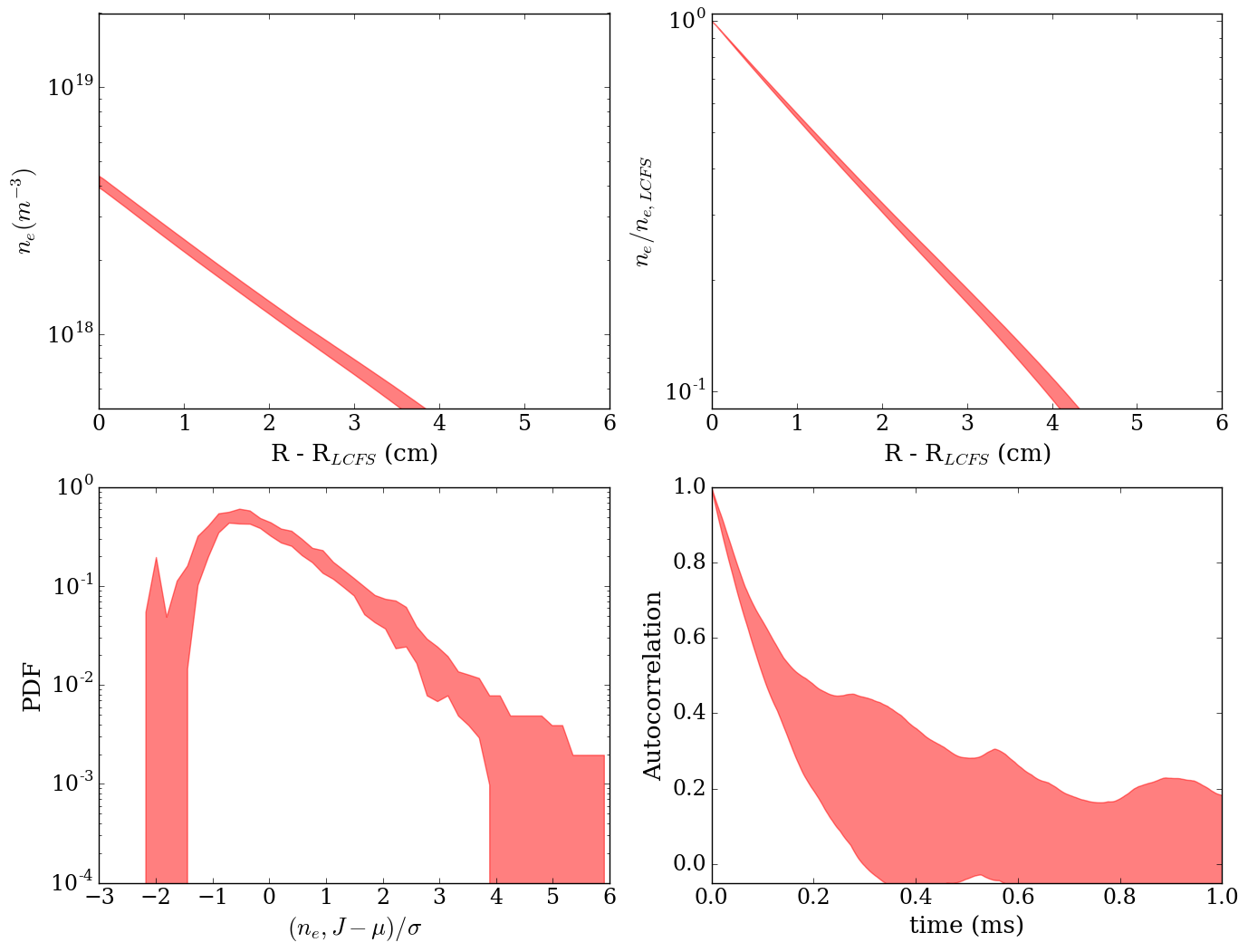}
    \caption{Measurement ranges from the convergence study carried out with a simulation time of 18ms. The absolute and relative profiles are shown in the upper row, whilst the PDF shape and autocorrelation function is shown in the bottom row. In each case, the range shown corresponds to the range produced in 10 identical simulations due to statistical variation.}
    \label{Fig:App_1}
\end{figure}

\begin{figure}[htbp]
    \centering
    \includegraphics[width=0.8\textwidth]{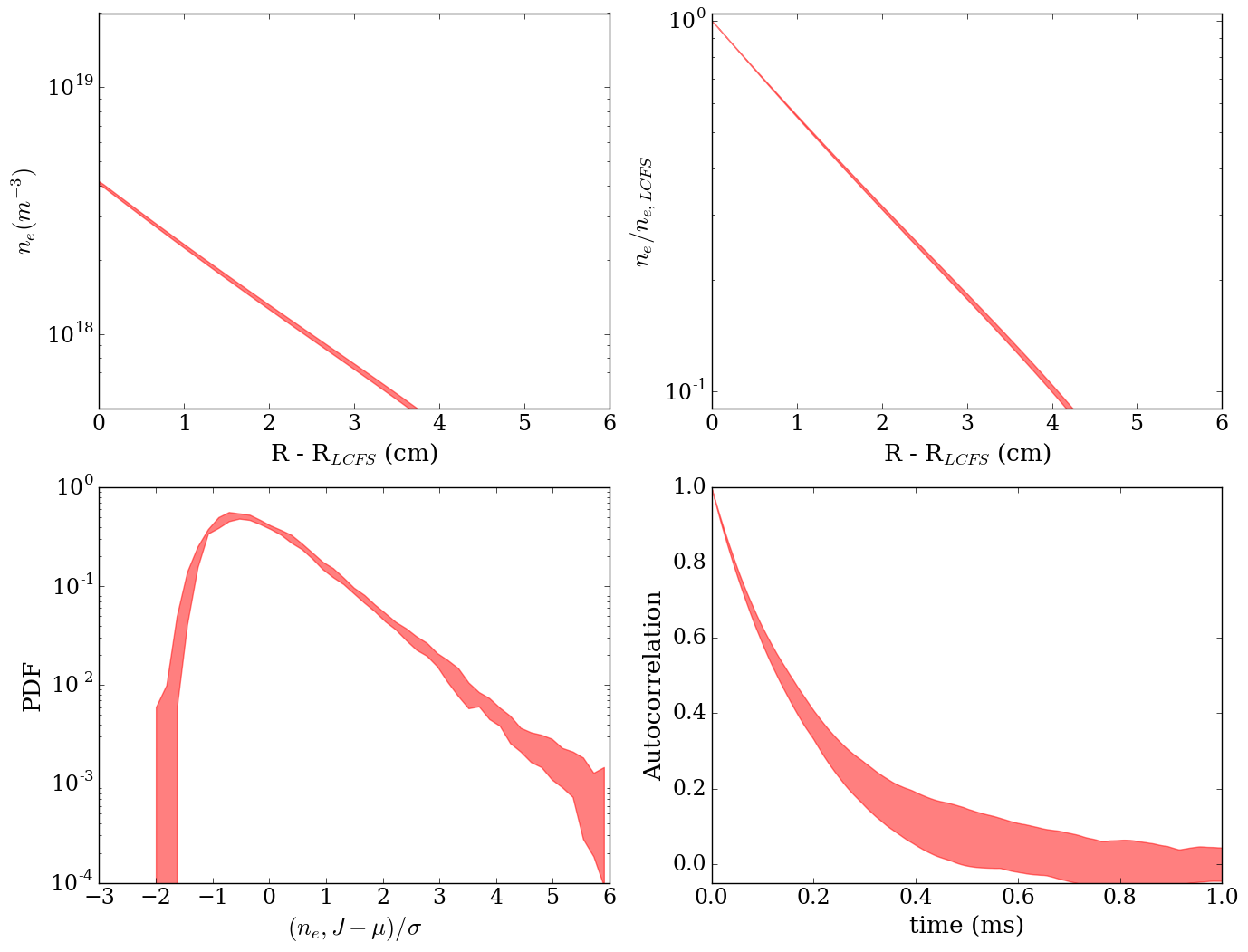}
    \caption{Measurement ranges from the convergence study carried out with a simulation time of 180ms. The absolute and relative profiles are shown in the upper row, whilst the PDF shape and autocorrelation function is shown in the bottom row. In each case, the range shown corresponds to the range produced in 10 identical simulations due to statistical variation.}
    \label{Fig:App_2}
\end{figure}

\begin{figure}[htbp]
    \centering
    \includegraphics[width=0.8\textwidth]{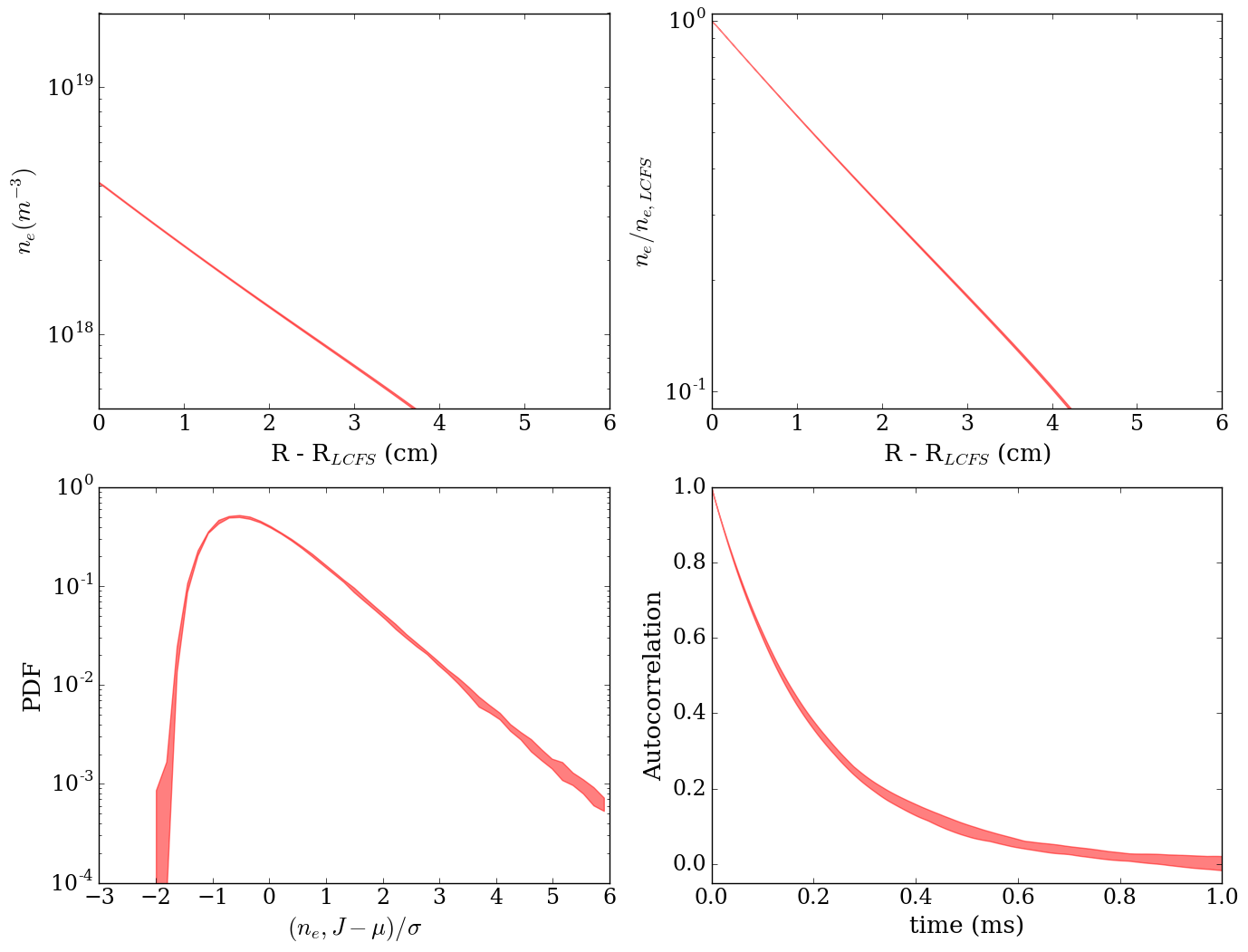}
    \caption{Measurement ranges from the convergence study carried out with a simulation time of 1.8s. The absolute and relative profiles are shown in the upper row, whilst the PDF shape and autocorrelation function is shown in the bottom row. In each case, the range shown corresponds to the range produced in 10 identical simulations due to statistical variation.}
    \label{Fig:App_3}
\end{figure}

In order to assess the statistical convergence of the measurements made in the stochastic model, simulations have been run with the simulation duration increased through two orders of magnitude. In each case, the simulations are repeated 10 times and the range of the measurements of the profile, PDF shape and autocorrelation functions are shown in figures \ref{Fig:App_1}, \ref{Fig:App_2} and \ref{Fig:App_3} respectively. 
This study shows that the profile can be considered converged for relatively short simulation times ($<18ms$) however long time-series ($>1.8s$) are required for decent statistical convergence of the PDF and the autocorrelation function. 

\end{document}